\documentclass[a4paper,aps,preprintnumbers,showpacs,twocolumn,superscriptaddress,nofootinbib,amsmath,amssymb]{revtex4}

\def\imo{i}

\newcommand{\ie}{{i.e.,}~}

\begin{document}
\title{Axisymmetric black holes allowing for separation of variables\\in the Klein-Gordon and Hamilton-Jacobi equations}
\author{R. A. Konoplya}\email{konoplya_roma@yahoo.com}
\affiliation{Theoretical Astrophysics, Eberhard-Karls University of T\"ubingen, T\"ubingen 72076, Germany}
\affiliation{Institute of Physics and Research Centre of Theoretical Physics and Astrophysics, Faculty of Philosophy and Science, Silesian University in Opava, CZ-746 01 Opava, Czech Republic}
\affiliation{Peoples Friendship University of Russia (RUDN University), 6 Miklukho-Maklaya Street, Moscow 117198, Russian Federation}
\author{Z. Stuchl\'ik}
\affiliation{Institute of Physics and Research Centre of Theoretical Physics and Astrophysics, Faculty of Philosophy and Science, Silesian University in Opava, CZ-746 01 Opava, Czech Republic}
\author{A. Zhidenko}\email{olexandr.zhydenko@ufabc.edu.br}
\affiliation{Centro de Matem\'atica, Computa\c{c}\~ao e Cogni\c{c}\~ao (CMCC), Universidade Federal do ABC (UFABC), Rua Aboli\c{c}\~ao, CEP: 09210-180, Santo Andr\'e, SP, Brazil}
\affiliation{Institute of Physics and Research Centre of Theoretical Physics and Astrophysics, Faculty of Philosophy and Science, Silesian University in Opava, CZ-746 01 Opava, Czech Republic}

\begin{abstract}
We determine the class of axisymmetric and asymptotically flat black-hole spacetimes for which the test Klein-Gordon and Hamilton-Jacobi equations allow for the separation of variables. The known Kerr, Kerr-Newman, Kerr-Sen and some other black-hole metrics in various theories of gravity are within the class of spacetimes described here. It is shown that although the black-hole metric in the Einstein-dilaton-Gauss-Bonnet theory does not allow for the separation of variables (at least in the considered coordinates), for a number of applications it can be effectively approximated by a metric within the above class. This gives us some hope that the class of spacetimes described here may be not only generic for the known solutions allowing for the separation of variables, but also a good approximation for a broader class of metrics, which does not admit such separation. Finally, the generic form of the axisymmetric metric is expanded in the radial direction in terms of the continued fractions and the connection with other black-hole parametrizations is discussed.
\end{abstract}
\pacs{04.50.Kd,04.70.-s}
\maketitle

\section{Introduction}

Recent progress in observations of black holes and their environment in the gravitational \cite{TheLIGOScientific:2016src} and electromagnetic \cite{Goddi:2016jrs} spectra made it important to develop a general formalism or parametrization allowing one to describe black holes in \emph{any} metric theory of gravity. The only general parametrization of this kind for axisymmetric and asymptotically flat black holes was suggested in \cite{Konoplya:2016jvv} and further tested in \cite{Ni:2016uik,Younsi:2016azx}. There, any axially symmetric and asymptotically flat black-hole metric can be represented in terms of a number of parameters which could be fixed through observations of the behavior of the black hole in both regions: near the event horizon (where the strong gravity regime matters) and far from the black hole (where the post-Newtonian expansion is implied). The usage of the continued fractions \cite{Rezzolla:2014mua} provided the strong hierarchy of the orders of parametrization and the quick convergence of the expansion used. Nevertheless, the price for being a truly general parametrization is a rather large number of parameters which must be fixed by the experimental data.

Therefore, here we will constrain the class of solutions and consider black-hole metrics written in the coordinates which guarantee the separation of variables in the Klein-Gordon and Hamilton-Jacobi equations. The other motivation for considering such special spacetimes is a simplification of the analysis of scalar fields and particles' motions, which makes it possible to give an analytical treatment for a number of problems, such as quasinormal modes, scattering and other wave phenomena, gravitational lensing, black hole shadows, etc.
After all, there is still no evidence that the black holes which we observe via gravitational waves or electromagnetic spectra of their environment have a symmetry which does not allow for the above separation of variables. The current uncertainty in the measurement of the mass and angular momentum of the resultant black hole does not allow to strongly constrain the black-hole geometry by observing the ringdown phase only \cite{Konoplya:2016pmh}.

However, here we shall not study such an issue as general conditions for the separability of the variables in both equations. The latter is intrinsically related to the symmetry of the background metric \cite{Bagrov:1974ru} and includes the search for the coordinates which allow the separability. The seminal work on the generic separability of the Klein-Gordon and Schr\"odinger equations in the background of a solution of the \emph{Einstein equations} (independently of the physical interpretation of this solution, and thus, not only, but including black holes) was done by Carter \cite{Carter:1968ks}\footnote{The electro-vacuum generalization in the Einstein theory, which includes the Kerr-NUT solution, was done in \cite{Dadhich:2001sz}.}. Being aimed at testing the strong gravity regime and alternatives of Einstein gravity, here we have a different task: to consider asymptotically flat and axisymmetric spacetimes describing a Kerr-like black hole, but in \emph{an arbitrary metric theory of gravity}. The Kerr-like black holes which we describe here have a number of features of the Kerr spacetime. They have the same quadrupole moment, possess a spherical horizon, mirror symmetry, etc.

Therefore, we will use a more practical approach. We know the way in which the separation of variables occurs for the well-known black-hole metrics, such as Kerr, Kerr-Newman, Kerr-Sen, etc. Here we generalize this procedure of the separation of variables in some sense. We show that, under the rather basic assumptions of asymptotic flatness, existence of a compact event horizon and a special way of separating the variables, the resultant class of black holes requires the spheroidal harmonics for the angular part of the dynamical equations, as it occurs for the Kerr solution. Such constrained spacetimes have three arbitrary functions of the radial coordinate which must be fixed to produce the known black-hole metrics (Kerr, Kerr-Newman, Kerr-Sen, and others) \emph{a priori} allowing for the above separation of variables.

Further we shall consider the rotating Einstein-dilaton-Gauss-Bonnet (EdGB) black-hole spacetime, which was obtained in \cite{Ayzenberg:2014aka} in the form of expansion in terms of \cite{Ayzenberg:2014aka} or slow rotation and coupling constants.  We shall show that the test scalar field in the background of this metric, written in the coordinates of \cite{Ayzenberg:2014aka} or \cite{Konoplya:2016jvv}, does not allow for the separation of variables. Nevertheless, we shall show that for a number of applications (such as the position and frequency of the innermost stable circular orbit, binding energy, etc.) the terms in the expansion violating the separation of variables can be safely neglected. This gives us some hope that the class of black-hole metrics described here and the associated parametrization derived can be used not only as an accurate description of Kerr-like black holes, but also as an effective approximation for much more general class of black-hole solutions.

The class of axisymmetric metrics allowing for the separation of variables in the Hamilton-Jacobi equations alone was considered recently in \cite{Johannsen:2015pca} and we shall show that the class of metrics considered in the present paper is a subclass of the metrics described in \cite{Johannsen:2015pca}. In the general case the Johannsen metric \cite{Johannsen:2015pca} does not allow for the separation of variables in the Klein-Gordon equations.

The paper is organized as follows. Sec.~\ref{sec:separability} is devoted to deduction of a general form of the metric allowing for the separation of variables in the Klein-Gordon and Hamilton-Jacobi equations. In this form there are three functions of the radial coordinates which must be fixed for any particular case. In Sec.~\ref{sec:Carter} we will relate the obtained metric with the Carter ansatz, while in  Sec.~\ref{sec:Johannsen} we will show that the axisymmetric metric of \cite{Johannsen:2015pca} does not allow for the separation of variables in the Klein-Gordon equation and includes our axisymmetric ansatz as a particular case. In Sec.~\ref{sec:examples} we write down the form of these three functions for various black-hole metrics, such as Kerr, Kerr-Newman, Kerr-Sen, etc. and show that the rotating Einstein-dilaton-Gauss-Bonnet black-hole spacetime does not belong to the considered class. In Sec.~\ref{sec:ISCO} we compute the binding energy of the EdGB black hole with and without terms that violate the separation of variables, while in Sec.~\ref{sec:parametrization} we do the same for the other characteristic variable: the frequency at the innermost stable circular orbit. Finally, in the Conclusion we discuss the obtained results and further perspectives of our approach.

\section{General form of the black-hole metric with separable Klein-Gordon and Hamilton-Jacobi equations}\label{sec:separability}
\subsection{Separation of variables in the Klein-Gordon equation}
Here we will imply that the spacetime under consideration is
\begin{enumerate}
\item\label{assump1} axially symmetric,
\item\label{assump2} asymptotically flat,
\item\label{assump3} possesses a compact event horizon.
\end{enumerate}

Having in mind the above, we will write the arbitrary axially symmetric metric as follows \cite{Konoplya:2016jvv}:
%
\begin{eqnarray}
ds^2 &=&
-\dfrac{N^2(r,\theta)-W^2(r,\theta)\sin^2\theta}{K^2(r,\theta)}dt^2
\nonumber \\&&
-2W(r,\theta)r\sin^2\theta dt \, d\phi
+K^2(r,\theta)r^2\sin^2\theta d\phi^2
\nonumber \\&&
+\Sigma(r,\theta)\left(\dfrac{B^2(r,\theta)}{N^2(r,\theta)}dr^2 +
r^2d\theta^2\right). \label{eq:initmetric}
\end{eqnarray}
The event horizon is determined by the equation $N(r, \theta) =0$.
Instead of angular variable $\theta$, we will use $y=\cos\theta$, so that the previous relation can be rewritten in the following way:
\begin{eqnarray}
ds^2 &=&
-\dfrac{N^2(r,y)-W^2(r,y)(1-y^2)}{K^2(r,y)}dt^2
\nonumber \\&&
-2W(r,y)r(1-y^2) dt \, d\phi+K^2(r,y)r^2(1-y^2) d\phi^2
\nonumber \\&&
+\Sigma(r,y)\left(\dfrac{B^2(r,y)}{N^2(r,y)}dr^2 +
r^2\dfrac{dy^2}{1-y^2}\right), \label{eq:metric}
\end{eqnarray}
We will require that the Klein-Gordon equation for a massive scalar field
\begin{equation}\label{boxphi}
(\Box-\mu^2)\Phi= 0
\end{equation}
not only allows for separation of variables in the chosen coordinates, but that this separation occurs in a similar fashion with the Kerr black hole. Namely, we imply that, after the extraction of a prefactor $N^p(r,y)B^q(r,y)$ from the generic wave-like function $\Phi(t,r,y,\phi)$, the equation for the remaining function $\Psi(r,y)$ is separable.
Taking into account the Killing vectors $\partial_t$ and $\partial_{\phi}$, this will lead to the following ansatz for the scalar field:
\begin{equation}\label{eq:ansatz}
\Phi(t,r,y,\phi)=e^{-\imo\omega t+\imo m\phi}N^p(r,y)B^q(r,y)\Psi(r,y).
\end{equation}

\begin{enumerate}\addtocounter{enumi}{3}
\item\label{assump4} The above method of separation of variables is our fourth assumption. Here one can certainly choose $p$ and $q$ to be zero, but we would like to see whether the nonzero values of $p$ and $q$ bring broader possibilities for the allowed forms of $N(r, y)$ and $B(r, y)$.
\end{enumerate}

Then, using (\ref{eq:ansatz}), the test field equation (\ref{boxphi}) takes the form
\begin{eqnarray}\label{eq:Klein-Gordon}
&e^{-\imo\omega t+\imo m\phi}\Biggr(P(r,y)\dfrac{\partial^2\Psi}{\partial y^2}+Q(r,y)\dfrac{\partial\Psi}{\partial y}
\\\nonumber&+T(r,y)\dfrac{\partial^2\Psi}{\partial r^2}+U(r,y)\dfrac{\partial\Psi}{\partial r}+ V(r,y, \omega, m) \Psi(r,y)\Biggr)=0,
\end{eqnarray}
where $P$, $Q$, $T$, and $U$ depend only on $N$, $B$, and their derivatives. The ``free term'' with the effective potential
$V(r,y, \omega, m)$ comes from $tt$, $t \phi$ and $\phi \phi$ derivatives in (\ref{boxphi}). Although the simplest case $p=q=0$ satisfies all the requirements, it would narrow the class of black-hole solutions considered here, allowing for the separation of variables.

The necessary conditions for the separability in the chosen variables ($r$ and $y$) are the following:

\begin{enumerate}\renewcommand\theenumi{\alph{enumi}} \renewcommand{\labelenumi}{\theenumi) }
\item The ratio
\begin{equation}\label{eq:yeq}
\dfrac{Q(r,y)}{P(r,y)}=\frac{4p}{N(r,y)}\dfrac{\partial N}{\partial y}+\frac{4q+1}{B(r,y)}\dfrac{\partial B}{\partial y}-\dfrac{2y}{1-y^2}
\end{equation}
must be a function of $y$ only.

\item The ratio
\begin{equation}\label{eq:req}
\dfrac{U(r,y)}{T(r,y)}=\frac{4p+2}{N(r,y)}\dfrac{\partial N}{\partial r}+\frac{4q-1}{B(r,y)}\dfrac{\partial B}{\partial r}+\dfrac{2}{r}
\end{equation}
must be a function of $r$ only.

\item\label{effpotcond} The effective potential $V(r,y, \omega, m)$  must also be representable in a form with separated variables in a special way which will be discussed below.
\end{enumerate}

We calculate derivatives of the right-hand side of (\ref{eq:yeq}) and (\ref{eq:req}) with respect to $r$ and $y$ and obtain two homogeneous equations, which for $p+q\neq-1/4$ have only the trivial solution
$$\dfrac{\partial^2}{\partial r\partial y}\ln(N)=\dfrac{\partial^2}{\partial r\partial y}\ln(B)=0,$$
implying that
$$N^2(r,y)=R_N(r)S_N(y),$$ 
and
$$B(r,y)=R_B(r)S_B(y).$$

Further we shall constrain ourselves to the asymptotically flat black holes, so that, since $$N^2(\infty,y)=B(\infty,y)=1,$$ we conclude that $S_N(y)=S_B(y)=1$.

When $p+q=-1/4$, the linear system (\ref{eq:yeq}) and (\ref{eq:req}) allows for a nontrivial solution
$$\dfrac{\partial^2}{\partial r\partial y}\ln\left(\dfrac{B}{N}\right)=0,$$
which we do not consider here, because the function $B$ vanishes at the event horizon ($N=0$).

Thus, we will continue with the metric which has the following form:
\begin{equation}\label{eq:NB}
N^2(r,y)=R_N(r),\qquad B(r,y)=R_B(r).
\end{equation}

Equation (\ref{eq:Klein-Gordon}) now takes the following form:
\begin{widetext}
\begin{eqnarray}\label{eq:Psi}
\dfrac{1}{1-y^2}\dfrac{\partial}{\partial y}(1-y^2)\dfrac{\partial\Psi}{\partial y}+\dfrac{r^2 R_N(r)}{R_B(r)^2}\dfrac{\partial^2\Psi}{\partial r^2}+\left(\dfrac{2r R_N(r)}{R_B(r)^2}+\dfrac{(2p+1)r^2 R_N'(r)}{R_B(r)^2}+\dfrac{(4q-1)r^2R_N(r)R_B'(r)}{R_B(r)^3}\right)\dfrac{\partial\Psi}{\partial r}
\\\nonumber
\Biggr(\omega^2\dfrac{r^2K^2(r,y)\Sigma(r,y)}{R_N(r)}-2m\omega\dfrac{r W(r,y)\Sigma(r,y)}{R_N(r)}
+\dfrac{m^2W^2(r,y)}{R_N(r)K^2(r,y)}-\dfrac{m^2\Sigma(r,y)}{(1-y^2)K^2(r,y)}-\mu^2r^2\Sigma(r,y)+H(r)\Biggr)\Psi&=&0,
\end{eqnarray}
\end{widetext}
where $H(r)$ depends on $R_N$, $R_B$, and their derivatives. For simplicity we here choose $p=q=0$, and thus $H(r)=0$. The choice of nonzero  $p$ and $q$, allowed by our assumption~(\ref{assump4}), evidently makes the calculations more involved, but does not lead to any broader class of the resultant metric functions $N(r, y)$ and $B(r, y)$.

In order to separate variables we require that the coefficient in front of $\Psi$ in (\ref{eq:Psi}) be sum of a function of $r$ and a function of $y$ for any given $\omega$, $m$, and $\mu$. This coefficient is an effective potential $V(r,y, \omega, m)$ up to the multiplier which depends on $r$ and $y$. Thus, here comes the separability of the effective potential which we mentioned in condition~(\ref{effpotcond}). Then, we imply that
\begin{eqnarray}\nonumber
W(r,y)&=&\dfrac{R_W(r)+S_W(y)R_N(r)/r}{\Sigma(r,y)},\\\nonumber
K^2(r,y)&=&\dfrac{R_K(r)+S_K(y)R_N(r)/r^2}{\Sigma(r,y)},\\\nonumber
\Sigma(r,y)&=&R_\Sigma(r)+S_\Sigma(y)/r^2.
\end{eqnarray}

Using again the asymptotical flatness, we observe that $W(r,y)={\cal O}(r^{-1})$ unless $S_W(y)=0$. Hence
\begin{equation}\label{eq:W}
W(r,y)=\dfrac{R_W(r)}{\Sigma(r,y)}.
\end{equation}
Further we will have to expand functions of $y$ around $y=0$, so that it is necessary to fix here the values of $S_W$, $S_K$ and $S_\Sigma$ at $y=0$. It is evident that as $R_W(r)$, $R_K(r)$ and $R_K(r)$ are some functions of $r$, which are not fixed, then the choice of constants $S_W(0)$, $S_K(0)$ and $S_\Sigma(0)$ will simply redefine the above functions of $r$, so that we choose $S_W(0)=S_K(0)=S_\Sigma(0)=0$ without loss of generality.

Let us now consider the condition on the factor in the two terms proportional to $m^2$ in eq. (\ref{eq:Psi}), which we denote as $F(r,y)$,
\begin{eqnarray}\label{eq:F}
&&F(r,y)\equiv\dfrac{W^2(r,y)}{R_N(r)K^2(r,y)}-\dfrac{\Sigma(r,y)}{(1-y^2)K^2(r,y)}
\\\nonumber &&=\dfrac{r^4 (1-y^2) R_W^2(r) - R_N(r) \left(r^2 R_\Sigma(r) + S_\Sigma(y)\right)^2}{r^2 (1 - y^2) R_N(r) \left(r^2 R_K(r) + R_N(r) S_K(y)\right)}.
\end{eqnarray}
First we find that, in the equatorial plane ($y=0$), one has
$$F(r,0)=\dfrac{R_W^2(r)}{R_K(r)R_N(r)}-\dfrac{R_\Sigma^2(r)}{R_K(r)}.$$
Since the asymptotic behavior
\begin{eqnarray}\nonumber
&&\lim_{r\to\infty}R_K(r)=\lim_{r\to\infty}R_N(r)=\lim_{r\to\infty}R_\Sigma(r)=1,
\\\nonumber
&&\lim_{r\to\infty}R_W(r)=0,
\end{eqnarray}
is fulfilled, then
\begin{equation}
F(r,0) \rightarrow -1, \quad \mbox{as} \quad r\to\infty
\end{equation}
In addition, we find that
$$\lim_{r\to\infty}F(r,y)=-1-\dfrac{y^2}{1-y^2}.$$

Since $F(r,y)$ is also a sum of a function of $r$ and a function of $y$, we conclude that
\begin{equation}\label{eq:Fe}
F(r,y)=\dfrac{R_W^2(r)}{R_K(r)R_N(r)}-\dfrac{R_\Sigma^2(r)}{R_K(r)}-\dfrac{y^2}{1-y^2}.
\end{equation}

Comparing the right-hand sides of (\ref{eq:F}) and (\ref{eq:Fe}), we obtain an equality, which must be satisfied for any $r$ and $y$
\begin{eqnarray}\label{eq:f}
\dfrac{r^4 (1-y^2) R_W^2(r) - R_N(r) \left(r^2 R_\Sigma(r) + S_\Sigma(y)\right)^2}{r^2 (1 - y^2) R_N(r) \left(r^2 R_K(r) + R_N(r) S_K(y)\right)}\\\nonumber=\dfrac{R_W^2(r)}{R_K(r)R_N(r)}-\dfrac{R_\Sigma^2(r)}{R_K(r)}-\dfrac{y^2}{1-y^2}.
\end{eqnarray}

Taking the limit $y\to1$ in the above relation (\ref{eq:f}), we find that
\begin{equation}\label{eq:RK}
R_K(r)=\left(R_\Sigma(r)+\dfrac{S_\Sigma(1)}{r^2}\right)^2-\dfrac{R_N(r)}{r^2}S_K(1).
\end{equation}

Substituting (\ref{eq:RK}) into (\ref{eq:f}) and expanding in terms of $r^{-1}$, we find that
\begin{equation}\label{eq:S}
S_\Sigma(y)=S_\Sigma(1)y^2,\qquad S_K(y)=S_K(1)y^2.
\end{equation}

Finally, using (\ref{eq:RK}) and (\ref{eq:S}), we reduce (\ref{eq:f}) to the following form
\begin{equation}\label{eq:RW}
R_W(r)=\dfrac{\pm\sqrt{S_K(1)}}{r}\left(R_\Sigma(r)\dfrac{S_\Sigma(1)}{S_K(1)}-R_N(r)+\dfrac{S_\Sigma^2(1)}{S_K(1)r^2}\right).
\end{equation}
It follows that $S_\Sigma(1)=S_K(1)>0$; otherwise $R_W(r)={\cal O}(r^{-1})$.

It is convenient now to introduce a rotation parameter $a\equiv\pm\sqrt{S_K(1)}$, so that
\begin{equation}\label{eq:Sa}
S_K(y)=S_\Sigma(y)=a^2y^2
\end{equation}
and
\begin{equation}\label{eq:RKa}
R_K(r)=\left(R_\Sigma(r)+\dfrac{a^2}{r^2}\right)^2-\dfrac{R_N(r)a^2}{r^2}.
\end{equation}

For convenience we introduce a new finite function $R_M(r)$, such that
\begin{equation}\label{eq:WM}
R_W(r)=\dfrac{aR_M(r)}{r^2}.
\end{equation}
Finally, \emph{three} arbitrary functions of $r$ remain: $R_M$, $R_B$, and $R_\Sigma$. The function $R_N(r)$, which defines the event horizon, is given
\begin{equation}\label{eq:NM}
R_N(r)=R_\Sigma(r)-\dfrac{R_M(r)}{r}+\frac{a^2}{r^2}.
\end{equation}

The general metric, for which we are able to separate variables, takes the following form:
\begin{subequations}\label{eq:gen}
\begin{eqnarray}
B(r,y)&=&R_B(r),\\
\Sigma(r,y)&=&R_\Sigma(r)+\dfrac{a^2y^2}{r^2},\label{eq:gensigma}\\
W(r,y)&=&\dfrac{aR_M(r)}{r^2\Sigma(r,y)},\label{eq:genW}\\
N^2(r,y)&=&R_\Sigma(r)-\dfrac{R_M(r)}{r}+\frac{a^2}{r^2},\\
K^2(r,y)&=&\dfrac{1}{\Sigma(r,y)}\Biggr(R_\Sigma^2(r)+R_\Sigma(r)\dfrac{a^2}{r^2}+\dfrac{a^2R_M(r)}{r^3}
\nonumber\\&&+\dfrac{a^2y^2}{r^2}\left(R_\Sigma(r)-\dfrac{R_M(r)}{r}+\frac{a^2}{r^2}\right)\Biggr)\label{eq:genK}\\
&=&\dfrac{1}{\Sigma(r,y)}\left(R_\Sigma^2(r)+R_\Sigma(r)\dfrac{a^2}{r^2}+\dfrac{a^2y^2}{r^2}N^2(r,y)\right)
\nonumber\\
&&+\dfrac{a W(r,y)}{r}.\nonumber
\end{eqnarray}
\end{subequations}

Thus, the metric coefficients depend on the tree functions of the radial coordinate $R_{B}(r)$, $R_{\Sigma}(r)$, and $R_{M}(r)$. However, we still have the freedom to choose the radial coordinate, so that only two of these three functions are independent. In the next section we will give examples of such choices. In particular, in a number of cases the choice $R_\Sigma(r)=1$ will be convenient, corresponding to the Boyer-Lindquist coordinates for the Kerr metric. The Kerr-Sen metric is usually given in the coordinates for which $R_B(r)=1$, corresponding to Schwarzschild-like coordinates in the nonrotating limit $a\to0$.

\subsection{Master equation for a scalar field}

The function $\Psi(r,y)=R(r)S(y)$, where the angular function satisfies the following equation:
\begin{eqnarray}
&&\Biggr(\dfrac{d}{dy}(1-y^2)\dfrac{d}{dy}+(\omega^2-\mu^2)a^2y^2\\\nonumber&&-\dfrac{m^2y^2}{1-y^2}-(m-a\omega)^2+\lambda\Biggr)S(y)=0,
\end{eqnarray}
where $\lambda$ is the separation constant, which can be found for any particular values of $\omega$, $a$, $\mu$ as spheroidal eigenvalues either numerically (see, for example, \cite{Kokkotas:2010zd} and reference therein) or, for instance, in the form of an expansion in terms of small $a \omega$ \cite{Starobinsky:1973aij}.
The radial equation has the form
\begin{eqnarray}
&&\Biggr(\dfrac{R_N(r)}{R_B(r)}\dfrac{d}{dr}\dfrac{r^2 R_N(r)}{R_B(r)}\dfrac{d}{dr}+\dfrac{(r^2\omega R_\Sigma(r) + a^2\omega-a m)^2}{r^2}\nonumber\\&&-R_N(r)(\lambda+\mu^2r^2R_\Sigma(r))\Biggr)R(r)=0.
\end{eqnarray}
Here
$$R_N(r)=N^2(r,y)=R_\Sigma(r)-\dfrac{R_M(r)}{r}+\frac{a^2}{r^2}.$$

By introducing a new function,
$$P(r)=\rho(r) R(r),\qquad \rho(r)=\sqrt{r^2R_\Sigma(r)+a^2},$$
and the tortoise coordinate as
$$dr_*=\dfrac{R_B(r)\rho^2(r)}{R_N(r)r^2}dr,$$
the radial equation is reduced to the wave-like form
\begin{equation}
\left(\dfrac{d^2}{dr_*^2}+\left(\omega-\dfrac{am}{\rho^2(r)}\right)^2-U(r)\right)P=0,
\end{equation}
with the effective potential
\begin{equation}
U(r)=\frac{R_N(r)r^2(\lambda+\mu^2r^2R_\Sigma(r))}{\rho^4(r)}+\dfrac{1}{\rho(r)}\dfrac{d^2\rho}{dr_*^2}.
\end{equation}

\subsection{Separation of variables in the Hamilton-Jacobi equation}
It is interesting to notice that the Hamilton-Jacobi equation,
\begin{equation}
g^{\mu\nu}\dfrac{\partial S}{\partial x^\mu}\dfrac{\partial S}{\partial x^\nu}+\mu^2=0,
\end{equation}
is also separable for the metrics (\ref{eq:gen}).

Indeed, after introducing the integrals of motion,
\begin{eqnarray}
\dfrac{\partial S}{\partial t}&=&E,\\
\dfrac{\partial S}{\partial \phi}&=&-L,
\end{eqnarray}
we obtain
\begin{eqnarray}
&&(1-y^2)\left(\dfrac{\partial S}{\partial y}\right)^2+\dfrac{y^2}{1-y^2}L^2-y^2a^2(E^2-\mu^2)
\\\nonumber&&
+\dfrac{r^2R_N(r)}{R_B^2(r)}\left(\dfrac{\partial S}{\partial r}\right)^2+\dfrac{R_\Sigma(r)}{R_N(r)}L^2-R_\Sigma(r)r^2(E^2-\mu^2)
\\\nonumber&&
-\dfrac{R_M(r)}{rR_N(r)}\left((L-aE)^2+R_\Sigma(r)r^2E^2\right)=0,
\end{eqnarray}
which evidently has separable variables.

In \cite{Carter:1973rla} it was shown that the separability of the Klein-Gordon equation usually implies the separability of the Hamilton-Jacobi equations as well and the latter implies also the existence of the additional (Carter) constant. This class of metrics belongs to type D according to the Petrov classification and describes compact objects whose multipole moments can be expressed in terms of mass and angular momentum. On the other hand, if the Hamilton-Jacobi equations are not separable, the  Klein-Gordon equations are usually not separable as well \cite{Carter:1973rla}. This agrees with our analysis, showing that for this particular method of separation of variables, the separability of the  Klein-Gordon equation constrains the allowed form of the metric more than the requirement of the separability of the Hamilton-Jacobi equations. Usually the existence of the Carter constant is associated with the Killing tensor of the second order, which should also be expected in our case. However, we do not study this issue and refer interested readers to the exhaustive review \cite{Frolov:2017kze}.

\subsection{Final remarks on the obtained class of metrics}

Summarizing this section, as a result of our assumptions~(\ref{assump1}-\ref{assump4}) we have obtained a class of metrics
given by Eqs.~(\ref{eq:metric})~and~(\ref{eq:gen}), which
\begin{itemize}
\item allow for separation of variables for the Klein-Gordon (with a massive term) and Hamilton-Jacobi equations;
\item have a spherical horizon;
\item have a well-defined angular momentum at infinity, corresponding to the asymptotic $g_{t \phi} = \frac{- 4 M a \sin^2 \theta}{r}$, as $r \rightarrow \infty$ (this behavior follows from the asymptotic flatness, which implies that $R_{M}=2M+{\cal O}(r^{-1})$); and
\item possess mirror symmetry relative to the equatorial plane, which follows from~(\ref{eq:Sa}).
\end{itemize}

One should note that, apart from the asymptotic flatness conditions, the radial coordinate is not fixed allowing for the freedom to choose one of the three radial functions in (\ref{eq:gen}).

For example, by defining the new radial coordinate through
$$dr'=R_B(r)dr,$$
one can transform the coordinates in order to have
\begin{equation}\label{eq:Cartercoord}
R_B(r')=1.
\end{equation}

Another choice of the new radial coordinate
$$r'=r\sqrt{R_\Sigma(r)},$$
leads to
\begin{equation}\label{eq:Paramcoord}
R_\Sigma(r')=1.
\end{equation}

In other words only two of the three functions, $R_M$, $R_B$, and $R_\Sigma$, are independent. The third one can be fixed by choosing the radial coordinate.

\section{Relation of the obtained metric with the earlier approaches}

Here we shall learn how the above-obtained black-hole metric is related to two similar approaches that exist in the literature. Namely, we shall discuss Carter's approach developed in  \cite{Carter:1973rla} and Johannsen's approach for the Hamilton-Jacobi equation \cite{Johannsen:2015pca}.

\subsection{Relation to the Carter ansatz}\label{sec:Carter}
In  \cite{Carter:1973rla} Carter suggested an axisymmetric generalization of the spherical spacetime allowing for the separation of variables. The resultant metric (not necessarily describing a black hole) was given in the following form (Eq.~(5.18)~in~\cite{Carter:1973rla})
\begin{eqnarray}\label{eq:Carter}
  &&ds^2=\left(C_yZ_r(r)-C_rZ_y(y)\right)\left(\frac{dr^2}{\Delta_r(r)}+\frac{dy^2}{\Delta_y(y)}\right)
  \\\nonumber&&+\frac{\Delta_y(y)[C_rdt-Z_r(r)d\phi]^2-\Delta_r(r)[C_ydt-Z_y(y)d\phi]^2}{C_yZ_r(r)-C_rZ_y(y)}.
\end{eqnarray}
Comparing (\ref{eq:Carter}) for large $r$ with the asymptotic expansion for a rotating body (see~(13) in~\cite{Konoplya:2016jvv})
\begin{eqnarray}\label{eq:asymptot}
d s^2 &\approx& - \left(1- \frac{2 M}{r}\right) d t^2- \frac{4 M a (1-y^2)}{r} d t\, d \phi \\\nonumber&&+ dr^2 + r^2\left(\frac{dy^2}{1-y^2}+(1-y^2) d\phi^2\right)\,,
\end{eqnarray}
we find that
\begin{subequations}\label{eq:flatcond}
\begin{eqnarray}
  \Delta_y(y) &=& 1-y^2,\\
  \Delta_r(r) &=& r^2\left(1-\frac{2M}{r}+{\cal O}\left(\frac{1}{r^2}\right)\right),\\
  Z_y(y)&=&a(1-y^2),\\
  Z_r(r) &=& r^2\left(1+{\cal O}\left(\frac{1}{r}\right)\right),\\
  C_y &=& 1, \\
  C_r &=& a.
\end{eqnarray}
\end{subequations}

The metric (\ref{eq:Carter}) with the conditions (\ref{eq:flatcond}) is a particular case of the metric (\ref{eq:initmetric}) with the functions (\ref{eq:gen}) given by
\begin{eqnarray}\nonumber
R_\Sigma(r)&=&\frac{Z_r(r)-a^2}{r^2},
\\
R_M(r)&=&\frac{Z_r(r)-\Delta_r(r)}{r},
\\
R_B(r)&=&1.\nonumber
\end{eqnarray}

In other words, the radial coordinate in Carter's paper is fixed as in (\ref{eq:Cartercoord}).

\subsection{Relation to the Johannsen deformation of Kerr}\label{sec:Johannsen}
In \cite{Johannsen:2015pca} Johannsen proposed a deformation of the Kerr black hole for which the Hamilton-Jacobi equation is separable. This weaker condition is satisfied by a more general class of the metrics, given by the following metric coefficients:
\begin{eqnarray}
g_{\theta \theta} &=& r^2 + f(r) + a^2\cos^2\theta, \nonumber \\
g_{rr} &=& \frac{g_{\theta \theta}}{\Delta(r) A_5(r)}, \qquad \Delta(r)\equiv r^2-2Mr+a^2,\nonumber \\
g_{tt} &=& -\frac{g_{\theta \theta}[\Delta(r)-a^2A_2(r)^2\sin^2\theta]}{[(r^2+a^2)A_1(r)-a^2A_2(r)\sin^2\theta]^2}, \label{eq:metric-Johannsen} \\
g_{t\phi} &=& \frac{a[\Delta(r)-(r^2+a^2)A_1(r)A_2(r)]g_{\theta \theta}\sin^2\theta}{[(r^2+a^2)A_1(r)-a^2A_2(r)\sin^2\theta]^2}, \nonumber \\
g_{\phi \phi} &=& \frac{g_{\theta \theta} \sin^2 \theta \left[(r^2 + a^2)^2 A_1(r)^2 - a^2 \Delta(r) \sin^2 \theta \right]}{[(r^2+a^2)A_1(r)-a^2A_2(r)\sin^2\theta]^2},
\nonumber
\end{eqnarray}

For the metric (\ref{eq:metric-Johannsen}) it is possible to show that the equation (\ref{eq:Klein-Gordon}) is not separable, unless the following condition is satisfied:
\begin{equation}\label{eq:separability}
  f(r)=(r^2+a^2)\left(\frac{A_1(r)}{A_2(r)}-1\right).
\end{equation}

When (\ref{eq:separability}) is fulfilled, then Eqs.~(\ref{eq:metric-Johannsen}) are reduced to (\ref{eq:initmetric}) with the functions (\ref{eq:gen}) given by
\begin{eqnarray}\nonumber
R_\Sigma(r)&=&\left(1+\frac{a^2}{r^2}\right)\frac{A_1(r)}{A_2(r)}-\frac{a^2}{r^2}=1+\frac{f(r)}{r^2},
\\
R_M(r)&=&\frac{(r^2+a^2)A_1(r)}{rA_2(r)}-\frac{r^2-2Mr+a^2}{rA_2(r)^2},
\\
R_B(r)&=&\frac{1}{A_2(r)\sqrt{A_5(r)}}.\nonumber
\end{eqnarray}
Thus, it can be seen that the class of metrics described here is a subclass of the metrics considered in \cite{Johannsen:2015pca}.

\section{Axisymmetric black holes}\label{sec:examples}
Here, in the first four subsections we shall show how various known black-hole metrics which allow for the above separation of variables can be represented within the obtained class of metric by fixing the functions $R_M(r)$, $R_\Sigma(r)$, and $R_B(r)$. The last subsection is devoted to the rotating black hole in the Einstein-dilaton-Gauss-Bonnet theory which, in the general case, does not allow for the separation of variables, but which, as it will be shown, for some applications can effectively be approximated by a metric allowing for the separation of variables.
\subsection{Kerr metric}
Choosing in Eqs.~(\ref{eq:gen})
$$R_M(r)=2M, \qquad R_\Sigma(r)=1, \qquad R_B(r)=1,$$
we obtain
\begin{eqnarray}
B(r,y)&=&1,\nonumber\\
\Sigma(r,y)&=&1+\dfrac{a^2y^2}{r^2},\nonumber\\
W(r,y)&=&\dfrac{2Ma}{r^2+a^2y^2},\\
N^2(r,y)&=&1-\dfrac{2M}{r}+\frac{a^2}{r^2},\nonumber\\
K^2(r,y)&=&\dfrac{r^2+a^2+2Ma^2/r+a^2y^2N^2(r,y)}{r^2+a^2y^2}\nonumber\\
&=&\dfrac{r^2+2a^2+a^4/r^2+a^2(y^2-1)N^2(r,y)}{r^2+a^2y^2}.\nonumber
\end{eqnarray}
Then (\ref{eq:initmetric}) reproduces the Kerr metric in the Boyer-Lindquist coordinates
\begin{eqnarray}\nonumber
 d s^2 &=& -\frac{r^{2}+a^2-2Mr}{r^{2}+a^2\cos^{2}\theta}\left(dt-a\sin^{2}\theta d\varphi\right)^{2}
 \\\label{eq:Kerr}&&
 +\frac{\sin^{2}\theta}{r^{2}+a^2\cos^{2}\theta}\left((r^2+a^2)d\varphi -a dt\right)^{2}
 \\\nonumber&&
 +\frac{r^{2}+a^2\cos^{2}\theta}{r^{2}+a^2-2Mr}dr^2 +(r^{2}+a^2\cos^{2}\theta)d\theta^{2}.
\end{eqnarray}

\subsection{Kerr-Newman metric}
The Kerr-Newman metric is obtained by choosing
$$R_M(r)=2M-\dfrac{Q^2}{r}, \qquad R_\Sigma(r)=1, \qquad R_B(r)=1,$$
so that
\begin{eqnarray}
B(r,y)&=&1,\nonumber\\
\Sigma(r,y)&=&1+\dfrac{a^2y^2}{r^2},\nonumber\\
W(r,y)&=&\dfrac{2Ma}{r^2+a^2y^2}-\dfrac{Q^2a}{r(r^2+a^2y^2)},\\
N^2(r,y)&=&1-\dfrac{2M}{r}+\frac{a^2+Q^2}{r^2},\nonumber\\
K^2(r,y)&=&\dfrac{r^2+a^2+2Ma^2/r-Q^2a^2/r^2+a^2y^2N^2(r,y)}{r^2+a^2y^2}\nonumber\\
&=&\dfrac{r^2+2a^2+a^4/r^2+a^2(y^2-1)N^2(r,y)}{r^2+a^2y^2}.\nonumber
\end{eqnarray}
Then (\ref{eq:initmetric}) reproduces the Kerr-Newman metric in the Boyer-Lindquist coordinates
\begin{eqnarray}\nonumber
 d s^2 &=& -\frac{r^{2}+a^2-2Mr+Q^2}{r^{2}+a^2\cos^{2}\theta}\left(dt-a\sin^{2}\theta d\varphi\right)^{2}
 \\\label{eq:Kerr-Newman}&&
 +\frac{\sin^{2}\theta}{r^{2}+a^2\cos^{2}\theta}\left((r^2+a^2)d\varphi -a dt\right)^{2}
 \\\nonumber&&
 +\frac{r^{2}+a^2\cos^{2}\theta}{r^{2}+a^2-2Mr+Q^2}dr^2 +(r^{2}+a^2\cos^{2}\theta)d\theta^{2}.
\end{eqnarray}

\subsection{Modified Kerr metric}
Choosing
$$R_M(r)=2M+\dfrac{\eta}{r^2}, \qquad R_\Sigma(r)=1, \qquad R_B(r)=1, $$
we obtain a deformation of the Kerr which does not change the post-Newtonian parameters
\begin{eqnarray}
B(r,y)&=&1,\nonumber\\
\Sigma(r,y)&=&1+\dfrac{a^2y^2}{r^2},\nonumber\\
W(r,y)&=&\dfrac{2Ma}{r^2+a^2y^2}+\dfrac{\eta a}{r^2(r^2+a^2y^2)},\\
N^2(r,y)&=&1-\dfrac{2M}{r}+\frac{a^2}{r^2}-\frac{\eta}{r^3},\nonumber\\
K^2(r,y)&=&\dfrac{r^2+a^2+2Ma^2/r+\eta a^2/r^3+a^2y^2N^2(r,y)}{r^2+a^2y^2}\nonumber\\
&=&\dfrac{r^2+2a^2+a^4/r^2+a^2(y^2-1)N^2(r,y)}{r^2+a^2y^2}.\nonumber
\end{eqnarray}

In this way in~(\ref{eq:initmetric}) we obtain the black-hole metric suggested in \cite{Konoplya:2016pmh}
\begin{eqnarray}\nonumber
 d s^2 &=& -\frac{r^{2}+a^2-2Mr-\eta/r}{r^{2}+a^2\cos^{2}\theta}\left(dt-a\sin^{2}\theta d\varphi\right)^{2}
 \\\label{eq:modified-Kerr}&&
 +\frac{\sin^{2}\theta}{r^{2}+a^2\cos^{2}\theta}\left((r^2+a^2)d\varphi -a dt\right)^{2}
 \\\nonumber&&
 +\frac{r^{2}+a^2\cos^{2}\theta}{r^{2}+a^2-2Mr-\eta/r}dr^2 +(r^{2}+a^2\cos^{2}\theta)d\theta^{2}.
\end{eqnarray}

\subsection{Kerr-Sen metric}
The Kerr-Sen metric is obtained by choosing
$$R_M(r)=2M, \qquad R_\Sigma(r)=1+\dfrac{2b}{r}, \qquad R_B(r)=1.$$
After introducing $\mu=M-b$ we obtain
\begin{eqnarray}
B(r,y)&=&1,\nonumber\\
\Sigma(r,y)&=&\dfrac{r^2+2br+a^2y^2}{r^2},\nonumber\\
W(r,y)&=&\dfrac{2Ma}{r^2+2br+a^2y^2}=\dfrac{2(\mu+b)a}{r^2+2br+a^2y^2},\\
N^2(r,y)&=&1+\dfrac{2b}{r}-\dfrac{2M}{r}+\frac{a^2}{r^2}=\dfrac{1-2\mu r+a^2}{r^2},\nonumber\\
K^2(r,y)&=&\dfrac{(r+2b)^2+a^2+2(M+b)a^2/r+a^2y^2N^2(r,y)}{r^2+2br+a^2y^2}\nonumber\\
&=&\dfrac{(r+2b+a^2/r)^2+a^2(y^2-1)N^2(r,y)}{r^2+2br+a^2y^2}.\nonumber
\end{eqnarray}

The metric~(\ref{eq:initmetric}) reads
\begin{eqnarray}\label{eq:Kerr-Sen}
 &&d s^2 = -\dfrac{r^{2}+a^2-2\mu r}{r^{2}+2br+a^2\cos^{2}\theta}\left(dt-a\sin^{2}\theta d\varphi\right)^{2}
 \\\nonumber&&+\dfrac{\sin^{2}\theta}{r^{2}+2br+a^2\cos^{2}\theta}\left((r^2+2br+a^2)d\varphi -a dt\right)^{2}
 \\\nonumber&&+\dfrac{r^{2}+2br+a^2\cos^{2}\theta}{r^{2}+a^2-2\mu r}dr^2 +(r^{2}+2br+a^2\cos^{2}\theta)d\theta^{2}.
\end{eqnarray}

\subsection{Black hole in the Einstein-dilaton-Gauss-Bonnet theory}

The metric for the axially symmetric black hole in the Einstein-dilaton-Gauss-Bonnet theory was found perturbatively in \cite{Ayzenberg:2014aka}. Here we shall use that metric rewritten in terms of the radial coordinate, such that (\ref{eq:gensigma}) is satisfied, and, consequently, expanded with respect to $y$ \cite{Konoplya:2016jvv}. This way, we take the black-hole metric in the form with $R_\Sigma(r)=1$,
\begin{widetext}
\begin{subequations}\label{EDGBmetric}
\begin{eqnarray}
\Sigma&=&1 + \frac{a^2}{r^2} y^2\,,\\
N^2 &=& 1 - \frac{2 M}{r} + \frac{a^2}{r^2} +
\zeta\left(\frac{M^3}{3 r^3} + \frac{26 M^4}{3 r^4} + \frac{22 M^5}{5
  r^5} + \frac{32 M^6}{5 r^6} - \frac{80 M^7}{3 r^7}\right)\\
&\phantom{=}& -\zeta\frac{a^2}{r^2}\Biggr(\frac{3267 M}{1750 r} +
\frac{5017M^2}{875 r^2} + \frac{136819 M^3}{18375 r^3} + \frac{35198
  M^4}{18375 r^4} - \frac{3818 M^5}{735 r^5} - \frac{4504 M^6}{245 r^6} +
\frac{16 M^7}{5 r^7}\Biggr) 
%
+ {\cal O}(a^3,\zeta^2,y^2)\,,
\nonumber\\
B &=& 1-\zeta\left(\frac{M^2}{2r^2}+\frac{4 M^3}{3 r^3}+\frac{7
  M^4}{r^4}+\frac{64 M^5}{5 r^5}+\frac{24 M^6}{r^6}\right)
\\
&\phantom{=}& +\zeta\frac{a^2}{r^2}\Biggr(\frac{1}{4}+\frac{2071 M}{875
  r}+\frac{2949 M^2}{350 r^2}+\frac{122862 M^3}{6125 r^3} + \frac{3014
  M^4}{105 r^4}+\frac{6396 M^5}{245 r^5}-\frac{48 M^6}{5 r^6}\Biggr)+{\cal O}(a^3,\zeta^2,y^2)\,,\nonumber
\end{eqnarray}
\begin{eqnarray}
W& = &\frac{2 M a }{r^2}\Biggr[1-\zeta\Biggr(\frac{3
  M^2}{10 r^2}+\frac{14 M^3}{3 r^3}+\frac{3 M^4}{r^4}+\frac{24 M^5}{5
  r^5} -\frac{40 M^6}{3 r^6}\Biggr)\Biggr]+{\cal
  O}(a^3,\zeta^2,y^4)\,,
\\
K^2 &=& 1+\frac{a^2}{r^2}+W\frac{a}{r}-\frac{2M a^2}{r^3}y^2
\\
&\phantom{=}& -\zeta\frac{a^2}{r^2}\Biggr(\frac{4463 M}{875 r} + \frac{2074
  M^2}{175 r^2} + \frac{127943 M^3}{6125 r^3} + \frac{4448 M^4}{525 r^4}
-\frac{2326 M^5}{245 r^5}-\frac{2792 M^6}{35 r^6} + \frac{16 M^7}{15
  r^7}\Biggr) y^2 
%
+ {\cal O}(a^4,\zeta^2,y^4)\,, \nonumber
\end{eqnarray}
\end{subequations}
where $\zeta \equiv 16\pi\alpha^2/\beta M^4$ is a small dimensionless parameter, $\alpha$ and $\beta$ are the Gauss-Bonnet and dilaton coupling constants, respectively.

It is easy to see that if one takes
\begin{subequations}\label{EDGBfuncdec}
\begin{eqnarray}
R_M(r) &=& 2 M -\zeta\left(\frac{M^3}{3 r^2} + \frac{26 M^4}{3 r^3} + \frac{22 M^5}{5
  r^4} + \frac{32 M^6}{5 r^5} - \frac{80 M^7}{3 r^6}\right)\\
&\phantom{=}& +\zeta\frac{a^2}{r^2}\Biggr(\frac{3267 M}{1750 r} +
\frac{5017M^2}{875 r^2} + \frac{136819 M^3}{18375 r^3} + \frac{35198
  M^4}{18375 r^4} - \frac{3818 M^5}{735 r^5} - \frac{4504 M^6}{245 r^6} +
\frac{16 M^7}{5 r^7}\Biggr)\,,
\nonumber
\end{eqnarray}
\begin{eqnarray}
R_B(r) &=& 1-\zeta\left(\frac{M^2}{2r^2}+\frac{4 M^3}{3 r^3}+\frac{7
  M^4}{r^4}+\frac{64 M^5}{5 r^5}+\frac{24 M^6}{r^6}\right)\\
&\phantom{=}& +\zeta\frac{a^2}{r^2}\Biggr(\frac{1}{4}+\frac{2071 M}{875
  r}+\frac{2949 M^2}{350 r^2}+\frac{122862 M^3}{6125 r^3}
 +\frac{3014 M^4}{105 r^4}+\frac{6396 M^5}{245 r^5}-\frac{48 M^6}{5 r^6}\Biggr)\,,\nonumber
\end{eqnarray}
\end{subequations}
\end{widetext}
then the functions $B$ and $N^2$, given by
\begin{subequations}\label{EDGBmetricdec}
\begin{eqnarray}
B(r,y)&=&R_B(r),\\
N^2(r,y)&=&1-\dfrac{R_M(r)}{r}+\frac{a^2}{r^2},
\end{eqnarray}
are reproduced precisely on the equatorial plane. Once the above choice is made and, in addition, if we require the separation of variables, the functions $W$ and $K^2$ must be defined as follows:
\begin{eqnarray}
W(r,y)&=&\dfrac{aR_M(r)}{r^2\Sigma(r,y)},\\
K^2(r,y)&=&\dfrac{1}{\Sigma(r,y)}\left(1+\dfrac{a^2}{r^2}+\dfrac{a^2y^2}{r^2}N^2(r,y)\right)\nonumber\\&&+\dfrac{a W(r,y)}{r}.
\end{eqnarray}
\end{subequations}
Then, the functions $W(r,y)$ and $K^2(r,y)$ get small corrections proportional to $a\zeta$ and $a^2\zeta$, respectively.

As a result we have obtained the metric, given by Eqs.~(\ref{EDGBmetricdec}), which allows for the separation of variables and can serve as an approximation of the EdGB black hole. In order to test this approximation in the equatorial plane in the next section we will calculate the frequency at the innermost stable circular orbit (ISCO) and the binding energy for the full black-hole metric~(\ref{EDGBmetric}) and its approximation derived here~(\ref{EDGBmetricdec}). These quantities do not depend on the function $B$.

\section{Binding energy and frequency at the innermost stable circular orbit}\label{sec:ISCO}

\subsection{Binding energy}

The binding energy is the amount of energy released by the particle going over from a given stable circular orbit located at $r_i$ to the ISCO at $r_{_{\rm ISCO}}$, \ie
\begin{equation}
\text{BE} = 1 - \frac{E(r_{_{\rm ISCO}})}{E(r_i)}\,,
\label{BE}
\end{equation}
where the initial circular orbit $r_i$ is normally considered to be at spatial infinity.
The binding energy is the simplest characteristic that may be associated with the amount of energy the accreting matter releases before falling onto the black-hole event horizon. It was calculated for various black holes in \cite{Konoplya:2006qr}.

Defining the four-momentum of a massive particle as
\begin{equation}
p^\alpha \equiv m \frac{dx^\alpha}{d s}\,,
\end{equation}
where $s$ is an invariant affine parameter, we also recall that in a
stationary, axisymmetric metric there are three integrals of motion which
can be related to the particle's rest mass $m$, to its energy $E = -p_t$,
and its angular momentum $L = p_\phi$. The normalization condition on the
four-momentum
\begin{equation}
p_\alpha p^\alpha = -m^2
\label{equation1}
\end{equation}
leads to the following relation in the equatorial plane
\begin{equation}
m^2 g_{rr}\left(\frac{dr}{d s}\right)^2 = V_{\rm eff}(r)\,,
\end{equation}
where the effective potential is defined as
\begin{equation}\label{effective-potential}
\begin{array}{l}
V_{\rm eff}(r) \equiv - \left( g^{tt}E^2 - 2g^{t\phi}E L + g^{\phi\phi}L^2 + m^2
  \right)\Biggr|_{\theta = \pi/2} =  \\
\dfrac{K^2(r,\pi/2)}{N^2(r,\pi/2)}
\left(E-\dfrac{W(r,\pi/2)}{K^2(r,\pi/2)}\dfrac{L}{r}\right)^2\!\!-\dfrac{L^2}{r^2
    K^2(r,\pi/2)} - m^2\,.
\end{array}
\end{equation}
The energy $E$ and momentum $L$ of a particle on a circular orbit at
radial position $r$ can then be determined from the requirements that
\begin{equation}
V_{\rm eff}(r)  = 0, \qquad V_{\rm eff}'(r) = 0\,,
\end{equation}
where $'$ indicates a derivative in the radial direction. Once the expressions for $L(r)$ and $E(r)$ have been obtained in this way, the position of the ISCO, $r_{_{\rm ISCO}}$, is computed from the additional condition $V_{\rm eff}''(r) = 0$.

Finally, we find the binding energy for a particle in the background given by the metric (\ref{EDGBmetric}),
\begin{eqnarray}\label{57}
\text{BE} &= 1 - \dfrac{2 \sqrt{2}}{3} + \dfrac{2233 \zeta}{131220 \sqrt{2}} \pm \dfrac{a}{18\sqrt{3}M} \pm \dfrac{50659 a \zeta}{1049760 \sqrt{3}M} \nonumber\\&+ \dfrac{5a^2}{162 \sqrt{2}M^2} + \dfrac{1361569247 a^2\zeta}{34720812000 \sqrt{2}M^2}+{\cal
  O}(a^3,\zeta^2),\nonumber\\
\end{eqnarray}
where $+$ and $-$ signs correspond to the co- and counter-rotating ISCO, respectively.

The particle's binding energy in the background~(\ref{EDGBmetricdec}) is given by,
\begin{eqnarray}\label{58}
\text{BE} &= 1 - \dfrac{2 \sqrt{2}}{3} + \dfrac{2233 \zeta}{131220 \sqrt{2}} \pm \dfrac{a}{18\sqrt{3}M} \pm \dfrac{51617 a \zeta}{1049760 \sqrt{3}M} \nonumber\\&+ \dfrac{5a^2}{162 \sqrt{2}M^2} + \dfrac{1432768697  a^2\zeta}{34720812000 \sqrt{2}M^2}+{\cal
  O}(a^3,\zeta^2).\nonumber\\
\end{eqnarray}
The difference between (\ref{57}) and (\ref{58}) is practically negligible, being approximately 2 orders smaller than the effect expressed in the coefficients proportional to $a \zeta$ and $a^2 \zeta$.

\subsection{ISCO frequencies}

We also calculate the corresponding ISCO frequencies, which are given by the following general formula
\begin{equation}
\Omega_{ISCO}=\dfrac{d\phi}{dt}\Biggr|_{r=r_{_{\rm ISCO}}}=\dfrac{d\phi}{ds}/\dfrac{dt}{ds}\Biggr|_{r=r_{_{\rm ISCO}}}.
\label{ISCO}
\end{equation}

For the metric (\ref{EDGBmetric}) we find (cf.~\cite{Jefremov:2015gza})
\begin{eqnarray}\label{60}
\Omega_{ISCO}M=\dfrac{1}{6\sqrt{6}}\left(1+\frac{32159 \zeta}{87480}\right)\pm \frac{11a}{216M} \pm \frac{282203a \zeta}{4723920M}
\nonumber\\+ \dfrac{a^2}{6\sqrt{6}M^2}\left(\frac{59}{108}+\frac{3550244443 \zeta}{3857868000}\right)+{\cal
  O}(a^3,\zeta^2)\,,\nonumber\\
\end{eqnarray}
while for the metric (\ref{EDGBmetricdec}) we obtain
\begin{eqnarray}\label{61}
\Omega_{ISCO}M=\dfrac{1}{6\sqrt{6}}\left(1+\frac{32159 \zeta}{87480}\right)\pm \frac{11a}{216M} \pm \frac{289106a \zeta}{4723920M}
\nonumber\\+ \dfrac{a^2}{6\sqrt{6}M^2}\left(\frac{59}{108}+\frac{3784470568 \zeta}{3857868000}\right)+{\cal
  O}(a^3,\zeta^2)\,.\nonumber\\
\end{eqnarray}

Again the difference between (\ref{60})~and~(\ref{61}) is approximately 2 orders smaller than the effect expressed in the coefficients proportional to $a \zeta$ and $a^2 \zeta$. Summarizing this section we conclude that the terms violating the separability of variables in the Klein-Gordon equation can be safely neglected in the Einstein-dilaton-Gauss-Bonnet rotating black-hole metric, at least when one is limited by estimations of such effects as the binding energy, position and frequency of the innermost stable circular orbit in the equatorial plane, etc.

\section{Parametrization}\label{sec:parametrization}

In \cite{Konoplya:2016jvv} the most general parametrization for an arbitrary axially symmetric and asymptotically flat black hole in any metric theory of gravity was suggested. In other words any axially symmetric black-hole metric can be expressed in terms of a number of parameters, in a similar fashion as the post-Newtonian parametrization, but in such a way that some of the parameters are fixed by the spacetime behavior near the event horizon and others by the behavior far from the black hole. Here, contrary to the most general parametrization, we consider the more narrow class of black-hole metrics with a guaranteed separation of variables in the Klein-Gordon and Hamilton-Jacobi equations. This class of black holes evidently can be described within the general parametrization as well and the requirement of separability must impose some constrains on the values of some parameters in the general parametrization. Here we shall study this issue.

\subsection{The general parametrization for rotating black holes}
Here we shall briefly relate the general parametrization of asymptotically flat and axisymmetric black holes developed in \cite{Konoplya:2016jvv}.  The metric components are functions of both the radial and the polar angular coordinates, forcing a double expansion to obtain a generic axisymmetric metric expression. The radial dependence is expressed in the form of a continued-fraction expansion in terms of a compactified radial coordinate. The dependence on $y$ has the form of a Taylor expansion in terms of $y$. These choices lead to a superior convergence in the radial direction and to an exact limit on the equatorial plane. Thus, the expansion in $y$ has the following general form
\begin{subequations}\label{y-expansion}
\begin{eqnarray}\label{eq:Sigmachoice}
\Sigma&=&1 + \frac{a^2}{r^2} y^2=1 + \frac{a^2}{r_0^2}(1-x)^2 y^2\,,\\
N^2 &=& xA_0(x)+\sum\limits_{i = 1}^{\infty}A_i(x)y^i\,, \\
B &=& 1+\sum\limits_{i = 0}^{\infty}B_i(x)y^i\,,\\
W &=& \sum\limits_{i = 0}^{\infty}\dfrac{W_i(x)y^i}{\Sigma}\,, \\
K^2&=& 1+\dfrac{aW}{r}+\sum\limits_{i = 0}^{\infty}\dfrac{K_i(x)y^i}{\Sigma}\,,
\end{eqnarray}
\end{subequations}
where we introduced the compact coordinate
$$x\equiv1-\dfrac{r_0}{r},$$
and $r_0$ is the horizon radius in the equatorial plane.

The asymptotic coefficients in
\begin{subequations}
\begin{eqnarray}
\label{bdef}
B_i(x) &=& b_{i0}(1-x)+{\tilde B}_i(x)(1-x)^2\,,\\
\nonumber \\
\label{wdef}
W_i(x) &=& w_{i0}(1-x)^2+{\tilde W}_i(x)(1-x)^3\,,\\
\nonumber \\
\label{kdef}
K_i(x) &=& k_{i0}(1-x)^2+{\tilde K}_i(x)(1-x)^3\,,\\
\nonumber \\
\label{a0def}
A_0(x) &=& 1-\epsilon_0(1-x)+(a_{00}-\epsilon_0+k_{00})(1-x)^2
\nonumber \\
&& \phantom{1} +{\tilde A}_0(x)(1-x)^3\,,\\
\nonumber \\
\label{aidef}
A_{i>0}(x) &=& K_i(x)+\epsilon_{i}(1-x)^2+a_{i0}(1-x)^3+\nonumber\\
&& \phantom{K_i(x)}+{\tilde A}_i(x)(1-x)^4\,,
\end{eqnarray}
\end{subequations}
are designed in such a way that they fit the required post-Newtonian behavior far from the black hole and the functions ${\tilde A}_i(x)$, ${\tilde B}_i(x)$, ${\tilde W}_i(x)$, ${\tilde K}_i(x)$ are fixed by the behavior of the spacetime near the event horizon:
\begin{subequations}
\begin{eqnarray}
{\tilde A}_i(x)& = &\dfrac{a_{i1}}{1+\dfrac{a_{i2}x}{1+\dfrac{a_{i3}x}{1+\ldots}}}\,,\\
\nonumber \\
{\tilde B}_i(x)& = &\dfrac{b_{i1}}{1+\dfrac{b_{i2}x}{1+\dfrac{b_{i3}x}{1+\ldots}}}\,,\\
\nonumber \\
{\tilde W}_i(x)& = &\dfrac{w_{i1}}{1+\dfrac{w_{i2}x}{1+\dfrac{w_{i3}x}{1+\ldots}}}\,,\\
\nonumber \\
{\tilde K}_i(x)& = &\dfrac{k_{i1}}{1+\dfrac{k_{i2}x}{1+\dfrac{k_{i3}x}{1+\ldots}}}\,.
\end{eqnarray}
\end{subequations}

The radial coordinate is chosen in such a way that
\begin{eqnarray}
k_{00}&=&\dfrac{a^2}{r_0^2},\\\nonumber
k_{01}&=&0.
\end{eqnarray}

Notice, that the expansion in the radial direction alone can be effectively used to numerically represent spherically symmetric black-hole solutions in a compact analytical form \cite{Kokkotas:2017ymc}.

\subsection{Constrains on the general parametrization for a black hole allowing for the separation of variables}
In order to satisfy the condition~(\ref{eq:Sigmachoice}), we fix the radial coordinate as (\ref{eq:Paramcoord}).
Comparing the general expansion (\ref{y-expansion}) with the metric (\ref{eq:gen}) we find that for our class of metrics one has
\begin{subequations}
\begin{eqnarray}
A_{i>0}(x)&=&0,\\
K_{i>2}(x)&=&0,\\
W_{i>0}(x)&=&0,\\
B_{i>0}(x)&=&0,
\end{eqnarray}
\end{subequations}
while, taking into account the radial dependence, we find that
\begin{subequations}
\begin{eqnarray}
R_B&=&b_{00}(1-x)+\dfrac{b_{01}(1-x)^2}{1+\dfrac{b_{02}x}{1+\dfrac{b_{03}x}{1+\ldots}}}\,,\\
R_M&=&r_0\Biggr(1+\frac{a^2}{r_0^2}(1-x)^2+\epsilon_0x
\\\nonumber&&-(a_{00}-\epsilon_0)(1-x)x-\dfrac{a_{01}(1-x)^2x}{1+\dfrac{a_{02}x}{1+\dfrac{a_{03}x}{1+\ldots}}}\Biggr)\,.
\end{eqnarray}
\end{subequations}
Asymptotically $R_B$ and $R_M$ are given as
\begin{subequations}
\begin{eqnarray}
R_B&=&b_{00}(1-x)+{\cal O}(1-x)^2\\\nonumber&=&(\gamma-1)\dfrac{M}{r_0}(1-x)+{\cal O}(1-x)^2,\\
\label{eq:RMas}
R_M&=&r_0\left(1+\epsilon_0-a_{00}(1-x)+{\cal O}(1-x)^2\right)\\\nonumber&=&2M\left(1+(\beta-\gamma)\dfrac{M}{r_0}(1-x)+{\cal O}(1-x)^2\right),
\end{eqnarray}
\end{subequations}
where $\beta$ and $\gamma$ are the PPN parameters and $M$ is the asymptotic mass (by definition of the asymptotic parameters $\epsilon_0$, $a_{00}$ and $b_{00}$). The coefficients $b_{01},b_{02},b_{03},\ldots$ and $a_{01},a_{02},a_{03},\ldots$ describe the near-horizon geometry.

All the other coefficients can be expressed in terms of $\epsilon_0$, $a_{00},a_{01},a_{02},a_{03},\ldots$ and $b_{00},b_{01},b_{02},b_{03},\ldots$.
From the relation (\ref{eq:genW})
\begin{equation}\label{eq:ww}
W_0(x)=\dfrac{aR_M}{r_0^2}(1-x)^2,
\end{equation}
by substituting (\ref{wdef})~and~(\ref{eq:RMas}) and comparing the asymptotics from the left- and right-hand sides, we find that
\begin{equation}\label{eq:wacons}
w_{00}=\dfrac{2Ma}{r_0^2}=\dfrac{a(1+\epsilon_0)}{r_0}.
\end{equation}
By expanding (\ref{eq:ww}) near the horizon ($x=0$), we find
\begin{eqnarray}\nonumber
w_{01}&=&\dfrac{a}{r_0}\left(\dfrac{a^2}{r_0^2}-\epsilon_0\right)=\dfrac{a}{r_0}\times\dfrac{r_0^2-2Mr_0+a^2}{r_0^2},
\\\nonumber
w_{02}&=&1+\dfrac{r_0^2(a_{00}+a_{01})}{a^2-r_0^2\epsilon_0}=1+\dfrac{r_0^2(a_{00}+a_{01})}{r_0^2-2Mr_0+a^2},
\\\label{eq:wcons}
w_{03}&=&-1-\dfrac{r_0^2a_{00}}{a^2-r_0^2\epsilon_0}\\\nonumber&&-\dfrac{r_0^4a_{01}(a_{00}+a_{01})}{(a^2-r_0^2\epsilon_0)(a^2+r_0^2(a_{00}+a_{01}-\epsilon_0))}\\\nonumber&&+\dfrac{r_0^2a_{01}a_{02}}{a^2+r_0^2(a_{00}+a_{01}-\epsilon_0)},\ldots
\end{eqnarray}

In a similar manner, from the relation (\ref{eq:genK})
\begin{equation}
K_2(x)=(A_0(x)x-1)\dfrac{a^2}{r_0^2}(1-x)^2,
\end{equation}
we find that
\begin{equation}\label{eq:kacons}
k_{20}=0,
\end{equation}
and also, from the near-horizon expansion, the expressions for $k_{21},k_{22},k_{23}\ldots$
\begin{eqnarray}\nonumber
k_{21}&=&-\dfrac{a^2}{r_0^2},\\
\label{eq:kcons}
k_{22}&=&-\dfrac{a^2}{r_0^2}+2\left(\dfrac{a^2}{r_0^2}-\epsilon_0\right)+a_{00}+a_{01}
\\\nonumber&=&-\dfrac{a^2}{r_0^2}+2\dfrac{r_0^2-2Mr_0+a^2}{r_0^2}+a_{00}+a_{01},\\\nonumber
k_{23}&=&1+\dfrac{r_0^2(a_{01}+a_{01}a_{02}+\epsilon_0)}{a^2+r_0^2(a_{00}+a_{01}-2\epsilon_0)}\\\nonumber&&-\dfrac{{a^2+r_0^2(a_{00}+a_{01}-2\epsilon_0)}}{r_0^2},\ldots
\end{eqnarray}

From
\begin{equation}
A_2(x)=0,
\end{equation}
we obtain
$$
\epsilon_{2}+a_{20}(1-x)+{\tilde A}_2(x)(1-x)^2=(1-A_0(x)x)\dfrac{a^2}{r_0^2}\,.
$$
Again, from the asymptotic expansion we find that
\begin{eqnarray}\label{eq:aacons}
\epsilon_2&=&0,\\\nonumber
a_{20}&=&\dfrac{a^2(1+\epsilon_0)}{r_0^2}=\dfrac{2Ma^2}{r_0^3},
\end{eqnarray}
implying that the quadrupole momentum is given as
\begin{equation}
Q=-\dfrac{a_{20}r_0^3+M(k_{00}+k_{20}r_0^2)}{3}=-Ma^2,
\end{equation}
obeying the same relation as for the Kerr black hole.

Finally, from the near-horizon expansion, we find
\begin{eqnarray}\nonumber
a_{21}&=&-\dfrac{a^2}{r_0^2}\epsilon_0=\dfrac{a^2}{r_0^2}-\dfrac{2Ma^2}{r_0^3}\,,\\
\label{eq:acons}
a_{22}&=&1-\dfrac{a^2+r_0^2(a_{00}+a_{01})}{r_0^2\epsilon_0}\,,\\\nonumber
a_{23}&=&\dfrac{a^2+r_0^2(a_{00}+a_{01}-\epsilon_0)}{r_0^2\epsilon_0}\\\nonumber&&+\dfrac{r_0^2a_{01}(1+a_{02})}{a^2+r_0^2(a_{00}+a_{01}-\epsilon_0)}\,\ldots.
\end{eqnarray}
Summarizing this section, we can notice that the system of expressions Eqs.~(\ref{eq:wacons}), (\ref{eq:wcons}), (\ref{eq:kacons}), (\ref{eq:kcons}), (\ref{eq:aacons}), and (\ref{eq:acons}), gives us all the constraints on the values of the parameters of the general parametrization when one is limited to black holes allowing for the above-discussed separation of variables.

\section{Conclusions}\label{sec:conclusions}

It is important to notice that the class of black holes considered here is not the most general class of black holes allowing for the separation of variables in the Klein-Gordon and Hamilton-Jacobi equations. Instead, we were interested in a more practical question of describing \emph{black holes with the Kerr-like symmetry}: the metrics considered here have a spherical event horizon and the same quadrupole momentum as the Kerr solution, resulting in the same spheroidal harmonics for the angular part. This class of black holes, however narrow it is, includes the known exact analytical black-hole solutions and many other black-hole metrics obtained by deformations of the Kerr metric or stipulated by some braneworld or cosmological scenarios etc. For example, in addition to the  Kerr, Kerr-Newman, Kerr-Sen and the deformed non-Kerr spacetimes considered here, our parametrization can be used to describe a rotating black projected on a brane \cite{Kanti:2006ua} or black holes generated by a quintessential field \cite{Schee:2016yzb}.

Our class of metrics is different from that considered by Johannsen \cite{Johannsen:2015pca} in two aspects. First, our metrics allow for the separation of variables not only in the Hamiton-Jacobi, but also in the Klein-Gordon equation. This way our metrics form a subclass of spacetimes  considered in \cite{Johannsen:2015pca}. Another aspect is related to the parametrization with respect to the radial coordinate. While  Johannsen's approach is based on the Taylor $1/r$ expansion, we used the continued-fraction approach and achieved thereby
the superior convergence. Within the framework of Johannsen's approach there remains the problem of the roughly equal contribution of lower and higher terms of the expansion (treated in \cite{Konoplya:2016jvv,Rezzolla:2014mua}), when the clear hierarchy of parameters is absent. Finally, we have obtained a much simpler class of axisymmetric black-hole spacetimes, which allows for a simpler analytic treatment for a number of problems.

In addition, we have shown that for some physical applications the considered class of spacetimes can be an effective approximation for black-hole metrics which do not allow for the separation of variables. Our work can be further improved in this direction by the analysis of the accuracy of the proposed approximation of the EdGB black hole outside the equatorial plane.

\acknowledgments{
R.~K. and A.~Z. acknowledge support of the ``Project for fostering collaboration in science, research and education'' funded by the Moravian-Silesian Region, Czech Republic and of the Research Centre for Theoretical Physics and Astrophysics, Faculty of Philosophy and Science of Sileasian University at Opava. The publication has been prepared with the support of the ``RUDN University Program 5-100''.
 A.~Z. was partially supported by Conselho Nacional de Desenvolvimento Cient\'ifico e Tecnol\'ogico (CNPq), Brazil.  Z. S. acknowledges  the Albert Einstein Centre for Gravitation and Astrophysics supported under the Czech Science Foundation (Grant No. 14-37086G). The authors would like to thank Eloy Ayon-Beato for useful comments.}

\end{document}